# Astrometric and photometric observations of six brightest trans-Neptunian objects at the Kyiv comet station

*A. Baransky*[1]*, *O. Lukina*[2]†, *S. Borysenko*[3]

[1] Astronomical Observatory of Taras Shevchenko National University of Kyiv, 3 Observatorna Str., 04053 Kyiv, Ukraine
[2] Taras Shevchenko National University of Kyiv, 4 Glushkova Ave., 03127, Kyiv, Ukraine
Main Astronomical Observatory of the NAS of Ukraine, 27 Akademika Zabolotnoho Str., 03143, Kyiv, Ukraine

In this work we focused on observations of six trans-Neptunian objects (TNOs) whose apparent magnitudes are brighter than $20^m$. We present the results of astrometric and photometric observations of (134340) Pluto, (136108) Haumea, (136472) Makemake, (136199) Eris, (90482) Orcus and (20000) Varuna obtained at the Kyiv comet station (Code MPC 585) in 2017-2019. For observations we used the 0.7-m (*f*/4) reflector AZT-8 with the FLI PL4710 CCD camera and filters of Johnson – Cousins photometric system. From our images we measured the objects' astrometric positions, calculated apparent magnitudes in the *BVRI* (mostly *R*) bands using the aperture photometry method, and found the absolute magnitudes together with the colour indices in several bands. Analysing our results, we investigate the limitation on the astrometry and photometry of faint objects with the 0.7-m telescope.

## INTRODUCTION

Trans-Neptunian objects (TNOs) are the Solar System bodies which orbit the Sun at a greater average distance than Neptune. Although the properties of the bright TNOs are well-defined, the peripheral region of the Solar System generally remains poorly studied. Ongoing astrometric and photometric monitoring of TNOs enables better determination of their orbits, detection of variations in surface properties and investigation of their orbital and photometric evolution. Also, TNOs are the indicators of possible gravitational influence from undiscovered bodies including the hypothetical Ninth Planet [1]. During the last decade, multiple objects were discovered and observed both from ground-based and space observatories as a part of the focused monitoring of single bodies and the large-scale surveys. In this work we have chosen six bright objects (Pluto, Haumea, Makemake, Eris, Orcus, Varuna) that are possible to observe with the equipment at the Kyiv comet station (MPC code 585) because of their predicted apparent magnitude of $14^m - 20^m$.

Despite the growing number of space and large ground-based telescopes, small telescopes with the aperture under 1-m are still widely used for astronomical observations, in particular for monitoring programs. One of many examples is the 0.7-m (*f*/4) reflector AZT-8, which is one of a few sources of unique astronomical observations in Ukraine. Working with such equipment, it is important to know as much as possible about its limitations and the quality of the results obtained on the edge of observational ability. Due to the small sizes and large distances to the Sun and the Earth of TNOs, the vast majority of them lie close to or beyond the limit of observation for AZT-8. In this paper we present the results of our first observations of several TNOs using 0.7-m telescope at the Kyiv comet station.

## TARGET DESCRIPTION

*Pluto* is a Kuiper belt object and a namesake for a group of TNOs in the 2:3 orbital resonance with Neptune, plutinos. It has a system of 5 satellites, and Charon is the only moon of Pluto that is massive enough to be in hydrostatic equilibrium, also causing the barycenter of the system to be outside Pluto. Beyond Charon, there are four much smaller moons, which are, in ascending order of distance from Pluto,

[1] abaransky@ukr.net
[2] oleksandra.lukina@gmail.com

Styx, Nix, Kerberos, and Hydra [32]. In 2015, the observations by the New Horizons spacecraft were conducted using a remote sensing package that included imaging instruments and a radio science investigation tool, as well as spectroscopic and other experiments. Using the data from the New Horizons Long-Range Reconnaissance Imager, the diameter of Pluto was found to be 2377 km [20], making Pluto the largest known TNO. Pluto's period of rotation is known to be 6.387 days [19]. It passed perihelion in 1989.

*Haumea* is a resonance Kuiper belt object in the weak 7:12 orbital resonance with Neptune. It has two small satellites, Hi'iaka and Namaka, with the compound mass of approximately 0.5 percent of the mass of the system [24]. The data from the 2017 stellar occultation indicate the presence of a narrow and dense ring with the radius close to the 3:1 mean-motion resonance with Haumea's spin period [21]. It is the object of an unusual elongated shape with the largest axis estimated to be at least 2322 km [21]. The possible reasons for Haumea's shape are the very short period of rotation, 3.915 hours [14] and mass concentrations towards the nucleus [21]. It passed aphelion in 1992.

*Makemake* is a classical Kuiper belt object close to the 7:13 orbital resonance with Neptune, and it has a single known satellite [22]. Makemake has a high orbital inclination and its orbital properties are generally close to the ones of Haumea (see Table 1). The diameter of Makemake was investigated during the 2013 stellar occultation and estimated to be 1434 km [5]. Visible and near-infrared spectral observations of Makemake showed that this object has an overall homogeneous surface [31]. Recent results of the long-term monitoring of Makemake suggested that it has an almost spherical shape and estimated its rotational period as 22.826 hours [12]. Makemake passed aphelion in 1992.

*Eris* is classified as a scattered disk object due to its high orbital eccentricity, and it has one known moon Dysnomia [6]. The orbit of Eris has the inclination of 44 degrees, which is much higher than the values of Kuiper belt objects. The results of the 2011 occultation estimated its diameter to be 2326 km, making Eris the second-largest TNO after Pluto. It also indicated the albedo of 0.96, one of the highest values in the Solar System [29]. The rotational period of Eris estimated from the space-based photometric observations is 1.08 days [26]. Eris passed aphelion around 1977.

*Orcus* is a Kuiper belt object in the 2:3 orbital resonance with Neptune which is classified as plutino. Vanth, the large satellite of Orcus, is approximately half of its diameter with the albedo approximately three times smaller. In 2018, the diameter of Orcus was calculated to be 910 km |6] , and the rotational period of this object was found to be 10.47 hours |6]. Orcus passed aphelion in 2019.

Uoruno is a classical Kuiper belt object. In 2019, it was proposed that changes in the rotational light-curve shape of Varuna may be caused by the presence of a large undiscovered satellite 9]. Multi-band thermal observations from the Herschel Space Observatory in 2013 proposed the mean diameter of Varuna equal to 668 km 15]. The rotational period of Varuna is calculated to be 6.344 hours [2], and it is a possible reason for its elongated shape. Varuna will approach aphelion in 2071.

OBSERVATIONS

In the period 2017-2019, the observations of selected trans-Neptunian objects were conducted by Alexander Baransky and Oleksandra Lukina at the Kyiv Comet Station (MPC code 585), Lisnyky, Ukraine (see Figure 1). The standard Johnson-Cousins broadband filters BVRI were used with the FLI PL 4710 CCD camera at the prime focus $F = 2828$ mm) of the 0.7—m (*f*/4) reflector AZT—8. The detector consists of the 1024 x 1024 array of 13 μm pixels, which corresponds to the scale of 0.947" per pixel.

The log of observations is listed in Table 3, where $N_i$ is the number of images per observational night; *Exp.* (s) is the exposition in seconds; r (AU) is the heliocentric distance; Δ (AU) is the geocentric distance; α (deg) is the phase angle of the object; SNR is the signal-to-noise ratio. Time of the observations presented in Table 3 is a mean of exposure.

Firstly, using the software SkyMap, The Sky, and JPL ephemerides, we planned the observations and prepared equatorial coordinates of the targets. Then we used the MaxIm DL program to cool down the camera and control the process of observation. We took series (from 5 to 30 per filter) of images in multiple filters during one session.

The limitation for astrometric analysis is the ability to see the target on the image. With the 0.7-m telescope it can be performed for faint objects with the apparent magnitude up to 20$^m$ in case if the large enough number of images is available and the addition tool is used. We managed to obtain the astrometric positions for all cases, except the observations of Eris in *V* and *B* bands (18.5$^m$ in *R* band) and the observation of Varuna in *V* band (19.8$^m$ in *R* band). As a conclusion, the observations of targets with the *R* magnitudes in the range 18$^m$—20$^m$ generally gave satisfactory results only when observed in the *R* band. The limitation of photometric analysis is mainly determined by the signal-to-noise ratio given in Table 3.

ASTROMETRIC AND PHOTOMETRIC ANALYSIS

For astrometric measurements, the Astrometrica 4 software was used with the Gaia DR2 and UCAC 4

star catalogues. When it was impossible to perform the data reduction with a single image, an image addition tool was used to increase the signal-to-noise ratio (SNR). We combined multiple images into one using the Stack images function in Astrometrica. The software uses the object's coordinates as input and takes into account the direction of motion and the angular velocity of the TNO. Then it sums up the pixel values from individual images. After the process of image addition, if it was needed, the astrometric data reduction was performed. As a result, we obtained tables with the coordinates and approximate apparent magnitudes of objects for each moment, which were sent to the Minor Planet Supplement.

The orbits of TNOs and (O – C) residuals differences between the observed coordinates and the coordinates calculated from the bigger number of observations – for both RA and Dec were determined using the Find Orb (version March 17, 2019) software, combining our own observations with data from the Minor Planet Center database over the last 2-4 years. In Table 4 we present the results of astrometric observations of TNOs, where $N$ – number of astrometric observations published in the Minor Planet Supplement, RA residual – (O – C) right ascension residual, Dec residual – (O – C) declination residual, $\pm\sigma"$ – rms deviation for each coordinate, References – number of the Minor Planet Supplement issue.

We sent the observations to the Minor Planet Center database with the maximum (O – C) residual of less than 0.5". In total, 157 precise astrometric observations of 6 TNOs, obtained during 7 observation nights, are published in the Minor Planet Center database and the Minor Planet Supplement. The (O – C) RA residual varies from −0.254 to 0.289, $\pm\sigma"$ RA – from 0.058 to 0.487. The (O – C) Dec residual varies from −0.042 to 0.325, $\pm\sigma"$ Dec — from 0.040 to 0.383.

For the apparent magnitudes calculations, we used the photometric method from [17]. In traditional CCD photometry, differential extinction is assumed to be negligible because the field of view, when imaging through a medium to long focal length telescope, is typically only several arc minutes (about 16' for the AZT-8 telescope with CCD FLI PL 4710). On the other hand, we used 5-11 solar analogue reference stars with $B - V$ about 0.5 – 0.8. Therefore, the differences in colours of reference stars and objects were small. Given the above, we used the simplified formula to estimate the apparent magnitude of the target in each filter:

$$m_{tar} = (m^{inst}_{tar} - m^{inst}_{ref}) + m_{ref} \quad (1)$$

are instrumental magnitudes of the target and the reference star respectively, obtained using the aperture photometry in the MaxIm DL 5 program, and $m_{ref}$ – apparent magnitude of the reference star which was taken from the APASS R9 catalogue [11] accessed via the Aladin v10.0 software.

Values of the $I$ and $R$ star magnitudes, which are not presented in the APASS catalogue, were derived from the APASS $V$ and the Sloan $g, r, i$ magnitudes using the following transformations suggested by Munari [18]:

$$I_C = i - 0.3645 - 0.0743 \times (g - i) + 0.0037 \times (g - i)^2 \quad (2)$$

$$R_C = r - 0.1712 - 0.0775 \times (V - i) - 0.0290 \times (V - i)^2 \quad (3)$$

Apparent magnitudes with the errors of relative photometry are given in Table 5. For the calculation of these errors, we used the formula from [4]:

$$\sigma_m = 1.0857/SNR$$

Other sources of the errors that are not represented in the confidence intervals include the errors of magnitude of reference stars, errors caused by using the simplified Eq. (1) and the system transformations from Eq. (2) and (3), as well as the APASS catalogue errors. We used reference stars with the magnitudes $13^m - 15^m$, so that the SNR was over 100 and the corresponding errors did not exceed $0.01^m$. Errors caused by the simplification of Eq. (1) are due to not taking into account the difference between stars' and targets' colour indices. For our observations these errors are in the range $0.01^m - 0.11^m$. The error caused by using the transformations of Eq. (2) and (3) is approximately equal to $0.018^m$ [18]. Errors of the APASS catalogue for the reference stars lie in the range of $0.01^m - 0.13^m$.

PHYSICAL PARAMETERS CALCULATION

We calculated colour index as the difference between the values of apparent magnitude in the corresponding filters: $V - R = V_{mag} - R_{mag}$, $B - V = B_{mag} - V_{mag}$. The results are given in Table 5 for each night of observation separately. The errors of colour measurements were calculated as, for example, for $B - V$ colour:

$$\sigma(B - V) = \sqrt{\sigma_B^2 + \sigma_V^2}$$

Absolute magnitude of the TNOs was calculated using the formula for absolute magnitude of the Solar System bodies [27]

$$H = m - 5\lg(r\Delta)$$

where m – apparent magnitude ($B$, $V$ or $R$), $r$ – distance from an object to the Sun, $\Delta$ – distance from an object to the Earth (AU). The original formula includes the phase angle dependence, but here we assume that it is negligible ($\alpha < 1.6°$), which may cause the underestimation of absolute magnitude of $0.1^m - 0.2^m$.

The results of calculations are presented in Table 2, where $H_V$ – mean values of the calculated absolute magnitudes with the corresponding errors. For Eris and Varuna, only absolute magnitudes in the $R$ filter were calculated directly from the observations. The $H_R$ results are $-1.34 \pm 0.03$ for Eris, $3.45 \pm 0.10$ for Varuna and $-0.27 \pm 0.02$ for Makemake. For the calculation of $H_V$ for Eris and Varuna, the $V$ magnitudes were derived from our $R$ magnitudes and the literature colour indices given in Table 2 as $V = (V - R) + R$. Approximate errors for the derived values were calculated as $\sigma(V_{der}) = \sqrt{\sigma^2_{(V-R)} + \sigma^2_R}$

## DISCUSSION AND CONCLUSIONS

We performed the astrometric and photometric observations of six brightest TNOs at the AZT-8 0.7-m telescope. For each of our targets, we derived the absolute magnitude $H_V$.

In Table 2 we compared the obtained absolute magnitudes with the literature data. Generally, our results correlate with the previously known ones, but the calculated absolute magnitudes are lower than the published ones. A possible explanation is the fact that the opposition effect was not taken into account. Also, the observations of TNOs performed with AZT-8 are short-term, so the absolute magnitude in our work represents only a small fraction of the orbit and may lead to differences with the previously obtained values.

The paper [12] included the new values of the absolute magnitudes of Makemake, $H_R = -0.388 \pm 0.02$ in the $R$ filter, and the value given in Table 2 in the $V$ filer. It is an example of the use of a 0.7-m telescope ($R$ filter) as a part of the instrumentation for the long-term photometric monitoring of a faint object, which is a general method of obtaining the absolute magnitude. Our result in the $R$ filter is $H_R = -0.27 \pm 0.02$, which is underestimated.

The averaged $V - R$ colour indices obtained in this work and the corresponding literature data are given in Table 2. The $V - R$ values calculated from different periods of the observations for Haumea and Makemake are scattered, while [30] and [12] discuss that these values do not considerably change. For Makemake, the $V - R$ result $0.41 \pm 0.02$ obtained in [12] is inside our range of values. For other objects, our $V - R$ colours do not show close agreement with the reference values in Table 2.

$B - V$ colours are given in Table 5. Our $B - V$ values are $0.93 \pm 0.04$ for Makemake and $0.45 \pm 0.03$ for Haumea. Our $B - V$ index for Makemake is in agreement with the result $0.91 \pm 0.03$ from [12] within the margin of errors, while the $B - V$ result for Haumea is different from the value $0.64 \pm 0.01$ presented in [30].

The combined colour observations of the Pluto – Charon system have not been performed for a long time since the space telescopes and the New Horizons mission became the new sources of data for Pluto and its satellites separately. The separate $B - V$ colours from the Hubble Space Telescope are 0.87 for Pluto and 0.71 for Charon [7]. In comparison with this data, our result of $0.59 \pm 0.01$ is much lower. Generally, our observations lead to the colours which are not in agreement with the literature data. The erroneous results are caused by the simplifications discussed above.

The aperture of the telescope has allowed to obtain the appropriate SNR for minimization of the estimation errors. But, for more detailed and accurate estimations, either the larger instrumental apertures or the long-term programs are needed.

Table 1: Orbital parameters (MPC, epoch 31 May 2020)

| Object | $q$ | $e$ | $a$ | $i$ | $M$ | Peri | Node |
|---|---|---|---|---|---|---|---|
| Pluto | 29.574 | 0.25283 | 39.861 | 17.097 | 42.950 | 115.340 | 110.297 |
| Haumea | 34.766 | 0.19490 | 43.182 | 28.214 | 217.770 | 238.782 | 122.162 |
| Makemake | 38.102 | 0.16128 | 45.429 | 28.984 | 165.521 | 294.827 | 79.619 |
| Eris | 38.275 | 0.43603 | 67.867 | 44.040 | 205.989 | 151.638 | 35.951 |
| Orcus | 30.280 | 0.22702 | 39.174 | 20.592 | 181.736 | 72.310 | 268.799 |
| Varuna | 40.317 | 0.05616 | 42.716 | 17.221 | 119.166 | 262.179 | 97.372 |

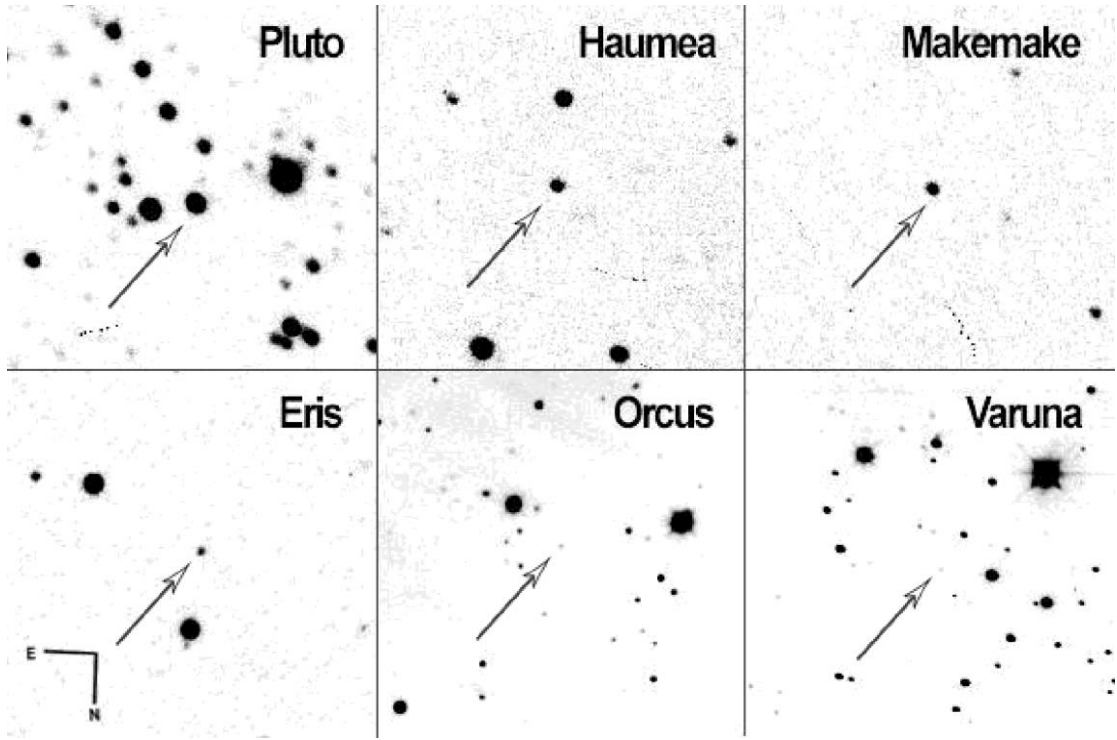

Fig. 1: R-band images of 6 observed TNOs: Pluto (2019-07-04), Haumea (2019-04-26), Makemake (2019-04-26), Eris (2018-10-05), Orcus (2019-03-02), Varuna (2019-03-02), centered. The FOV is 6' x 6', Kyiv comet station (585).

Table 2: Comparison of absolute magnitudes and $(V-R)$ colour indices with the literature data

| Object | $H$ | $H^{ref}$ | | $(V-R)$ | $(V-R)^{ref}$ | |
|---|---|---|---|---|---|---|
| Pluto | –0.40±0.01 | 0.648±0.01 | [25] | 0.72±0.01 | 0.60±0.02 | [8] |
| Haumea | 0.74±0.12 | 0.428±0.011 | [23] | 0.53±0.14 | 0.33±0.01 | [30] |
| Makemake | 0.18±0.13 | 0.049±0.020 | [12] | 0.45±0.11 | 0.41±0.02 | [12] |
| Eris | –0.89±0.03 | –1.12±0.03 | [28] | | 0.45±0.01 | [30] |
| Orcus | 2.43±0.09 | 2.31±0.03 | [10] | 0.85±0.10 | 0.37±0.04 | [3] |
| Varuna | 4.09±0.10 | 3.760±0.035 | [15] | | 0.64±0.02 | [13] |

Table 3: Observational details

| Object | Date | Time, UT | Filter | $N_i$ | Exp.(s) | r (AU) | Δ (AU) | α (deg) | SNR |
|---|---|---|---|---|---|---|---|---|---|
| Pluto | 2017-09-14 | 18:11:00 | R | 7 | 30 | 33.406 | 32.972 | 1.563 | 293 |
| | 2019-07-04 | 00:01:08 | R | 6 | 60 | 33.832 | 32.831 | 0.317 | 76 |
| | 2019-07-04 | 23:14:55 | B | 6 | 60 | 33.832 | 32.829 | 0.289 | 116 |
| | | 23:14:48 | V | 5 | 60 | | | | 158 |
| | | 23:12:48 | R | 6 | 60 | | | | 476 |
| Haumea | 2018-10-13 | 17:13:40 | R | 16 | 60 | 50.462 | 51.342 | 0.531 | 20 |
| | 2019-03-03 | 02:42:12 | B | 9 | 90 | 50.431 | 49.802 | 0.879 | 17 |
| | | 02:46:18 | V | 10 | 60 | | | | 28 |
| | | 02:50:13 | R | 9 | 60 | | | | 47 |
| | | 02:41:45 | I | 11 | 30 | | | | 21 |
| | 2019-04-26 | 22:39:56 | B | 16 | 60 | 50.419 | 49.549 | 0.577 | 45 |
| | | 22:31:48 | V | 16 | 60 | | | | 65 |
| | | 22:39:05 | R | 15 | 60 | | | | 90 |
| | 2019-07-03 | 20:42:06 | R | 21 | 60 | 50.404 | 50.192 | 1.131 | 107 |
| | | 20:43:09 | V | 21 | 60 | | | | 55 |
| Makemake | 2019-04-26 | 23:47:56 | B | 10 | 80 | 52.551 | 51.798 | 0.730 | 35 |
| | | 23:49:08 | V | 10 | 60 | | | | 48 |
| | | 23:48:27 | R | 9 | 60 | | | | 70 |
| | 2019-07-03 | 21:33:10 | V | 14 | 60 | 52.556 | 52.668 | 1.100 | 34 |
| | | 21:30:10 | R | 14 | 60 | | | | 73 |
| Eris | | 23:21:40 | R | 18 | 60 | | | | 37 |
| Orcus | 2019-03-02 | 23:52:01 | R | 27 | 60 | 48.068 | 47.136 | 0.405 | 21 |
| | 2019-03-03 | 00:22:16 | V | 29 | 60 | | | | 12 |
| Varuna | 2019-03-02 | 20:33:28 | R | 27 | 60 | 43.929 | 43.182 | 0.851 | 11 |

Table 4: The astrometric observations of TNOs and their accuracy (the average for each observation night)

| Object | Date | N | RA residual | ±σ″ | Decl. residual | ±σ″ | References |
|---|---|---|---|---|---|---|---|
| (20000) Varuna | 2019-03-02 | 15 | –0.097 | 0.260 | 0.261 | 0.243 | MPS 1166408 |
| (90482) Orcus | 2019-03-02 | 16 | –0.254 | 0.118 | 0.112 | 0.063 | MPS 1168275 |
| (134340) Pluto | 2017-09-14 | 6 | 0.176 | 0.177 | –0.008 | 0.277 | MPS 821439 |
| (134340) Pluto | 2019-07-03 | 4 | 0.023 | 0.090 | 0.166 | 0.166 | MPS 1169039 |
| (134340) Pluto | 2019-07-04 | 14 | –0.153 | 0.104 | 0.231 | 0.165 | MPS 1169039 |
| (136108) Haumea | 2018-10-13 | 5 | 0.289 | 0.487 | 0.308 | 0.383 | MPS 946328 |
| (136108) Haumea | 2019-03-03 | 15 | –0.000 | 0.203 | 0.325 | 0.145 | MPS 1169070 |
| (136108) Haumea | 2019-04-26 | 34 | 0.123 | 0.214 | –0.010 | 0.282 | MPS 1008259 |
| (136108) Haumea | 2019-07-03 | 8 | 0.059 | 0.075 | 0.304 | 0.040 | MPS 1169070 |
| (136199) Eris | 2018-10-05 | 13 | 0.103 | 0.058 | –0.042 | 0.148 | MPS 945212 |
| (136472) Makemake | 2019-04-26 | 20 | 0.046 | 0.158 | –0.041 | 0.192 | MPS 1008268 |
| (136472) Makemake | 2019-07-03 | 7 | 0.058 | 0.104 | 0.234 | 0.133 | MPS 1169075 |

Table 5: Results of photometric measurements

| Object | Date | $B$ (mag) | $V$ (mag) | $R$ (mag) | $I$ (mag) | $B-V$ (mag) | $V-R$ (mag) |
|---|---|---|---|---|---|---|---|
| Pluto | 2017-09-14 | | | 14.09±0.01 | | | |
| | 2019-07-03 | | | 14.06±0.01 | | | |
| | 2019-07-04 | 15.35±0.01 | 14.76±0.01 | 14.04±0.01 | | 0.59±0.01 | 0.72±0.01 |
| Haumea | 2018-10-13 | | | 16.92±0.05 | | | |
| | 2019-03-02 | 18.06±0.06 | 17.64±0.04 | 17.21±0.02 | 15.96±0.05 | 0.42±0.07 | 0.43±0.04 |
| | 2019-04-26 | 18.39±0.02 | 17.91±0.02 | 17.24±0.01 | | 0.48±0.03 | 0.67±0.02 |
| | 2019-07-03 | | 17.56±0.02 | 17.17±0.01 | | | 0.39±0.02 |
| Makemake | 2019-04-26 | 18.13±0.03 | 17.20±0.03 | 16.87±0.02 | | 0.93±0.04 | 0.33±0.04 |
| | 2019-07-03 | | 17.44±0.03 | 16.88±0.01 | | | 0.56±0.03 |
| Eris | 2018-10-05 | | | 18.47±0.03 | | | |
| Orcus | 2019-03-02 | | 19.214±0.09 | 18.364±0.05 | | | 0.85±0.10 |
| Varuna | 2019-03-02 | | | 19.84±0.10 | | | |